\documentclass[fleqn,usenatbib]{mnras}

\usepackage{newtxtext,newtxmath}

\usepackage[T1]{fontenc}

\DeclareRobustCommand{\VAN}[3]{#2}
\let\VANthebibliography\thebibliography
\def\thebibliography{\DeclareRobustCommand{\VAN}[3]{##3}\VANthebibliography}

\usepackage{graphicx}	
\usepackage{amsmath}	
\usepackage{eso-pic}
\usepackage{csvsimple}  
\usepackage{dirtytalk}  
\hypersetup{
	colorlinks=true,       
	linkcolor=blue,        
	citecolor=blue,        
	filecolor=magenta,     
	urlcolor=blue         
}
\usepackage{booktabs}   
\usepackage[switch]{lineno}

\usepackage{siunitx}
\DeclareSIUnit \h {\ensuremath{\mathit{h}}}
\DeclareSIUnit \pc {pc}

\title[Constraining small scale power with DES-SN5YR]{It's not $\sigma_8$ : constraining the non-linear matter power spectrum with the Dark Energy Survey Year-5 supernova sample}

 \AddToShipoutPictureBG*{%
  \AtPageUpperLeft{%
    \hspace{\paperwidth}%
    \raisebox{-\baselineskip}{%
      \makebox[0pt][r]{DES-2024-0863, \;\; FERMILAB-PUB-24-0769-V \;\;}
}}}%

\author[P. Shah et. al.]{
\parbox{\textwidth}{
\Large
P.~Shah,$^{1}$\thanks{E-mail: paul.shah.19@ucl.ac.uk}
T.~M.~Davis,$^{2}$
M.~Vincenzi,$^{3}$
P.~Armstrong,$^{4}$
D.~Brout,$^{3,5}$
R.~Camilleri,$^{2}$
L.~Galbany,$^{6,7}$
M.~S.~S.~Gill,$^{8}$
D.~Huterer,$^{9}$
N.~Jeffrey,$^{1}$
O.~Lahav,$^{1}$
J.~Lee,$^{10}$
C.~Lidman,$^{11,4}$
A.~M\"oller,$^{12}$
M.~Sullivan,$^{13}$
L.~Whiteway,$^{1}$
P.~Wiseman,$^{13}$
S.~Allam,$^{14}$
M.~Aguena,$^{15}$
J.~Annis,$^{14}$
J.~Blazek,$^{16}$
D.~Brooks,$^{1}$
A.~Carnero~Rosell,$^{17,15,18}$
J.~Carretero,$^{19}$
C.~Conselice,$^{20,21}$
L.~N.~da Costa,$^{15}$
M.~E.~S.~Pereira,$^{22}$
S.~Desai,$^{23}$
H.~T.~Diehl,$^{14}$
P.~Doel,$^{1}$
S.~Everett,$^{24}$
I.~Ferrero,$^{25}$
B.~Flaugher,$^{14}$
J.~Frieman,$^{14,26}$
J.~Garc\'ia-Bellido,$^{27}$
E.~Gaztanaga,$^{6,28,7}$
G.~Giannini,$^{19,26}$
D.~Gruen,$^{29}$
R.~A.~Gruendl,$^{30,31}$
G.~Gutierrez,$^{14}$
S.~R.~Hinton,$^{2}$
D.~L.~Hollowood,$^{32}$
K.~Honscheid,$^{33,34}$
D.~J.~James,$^{5}$
S.~Lee,$^{35}$
J.~L.~Marshall,$^{36}$
J. Mena-Fern{\'a}ndez,$^{37}$
R.~Miquel,$^{38,19}$
A.~Palmese,$^{39}$
A.~Pieres,$^{15,40}$
A.~A.~Plazas~Malag\'on,$^{41,8}$
A.~Porredon,$^{42,43}$
S.~Samuroff,$^{16,19}$
E.~Sanchez,$^{42}$
I.~Sevilla-Noarbe,$^{42}$
M.~Smith,$^{44}$
E.~Suchyta,$^{45}$
M.~E.~C.~Swanson,$^{30}$
G.~Tarle,$^{9}$
D.~L.~Tucker,$^{14}$
and N.~Weaverdyck$^{46,47}$
\begin{center} (DES Collaboration) \end{center}
}
\vspace{0.4cm}
\\
\textit{Affiliations are listed at the end of the paper}
}

\date{Accepted XXX. Received YYY; in original form ZZZ}

\pubyear{2024}

\begin{document}
\label{firstpage}
\pagerange{\pageref{firstpage}--\pageref{lastpage}}
\maketitle


\begin{abstract}
The weak gravitational lensing magnification of Type Ia supernovae (SNe Ia) is sensitive to the matter power spectrum on scales $k>1 h$ Mpc$^{-1}$, making it unwise to interpret SNe Ia lensing in terms of power on linear scales. We compute the probability density function of SNe Ia magnification as a function of standard cosmological parameters, plus an empirical parameter $A_{\rm mod}$ which describes the suppression or enhancement of matter power on non-linear scales compared to a cold dark matter only model. While baryons are expected to enhance power on the scales relevant to SN Ia lensing, other physics such as neutrino masses or non-standard dark matter may suppress power.  Using the Dark Energy Survey Year-5 sample, we find $A_{\rm mod} = 0.77^{+0.69}_{-0.40}$ (68\% credible interval around the median). Although the median is consistent with unity there are hints of power suppression, with $A_{\rm mod} < 1.09$ at 68\% credibility. 
\end{abstract}





\section{Introduction}
\label{sec:intro}
The Cosmic Microwave Background (CMB) together with the expansion history of the universe at redshifts $z<2$, as resolved by SNe Ia or Baryon Acoustic Oscillations (BAO), suggest the universe is geometrically flat, and in the main consistent with cold dark matter (CDM) and a cosmological constant $\Lambda$ as the dominant energy components at late times. While some hints have arisen recently of an evolving dark energy component \citep{Rubin2023,SNKeyPaper, DESI2024, Camilleri2024} or physics that mimics it, if the universe is not $\Lambda$CDM it is at least very close to it in expansion history.
\par
Nevertheless, measurements of the clustering of matter in $\Lambda$CDM differ between those measured in the late universe and those predicted from the spectrum of $\mathcal{O} (10^{-5})$ CMB temperature and polarisation fluctuations projected to the present day using the inferred expansion history and standard gravity \citep[see][and references therein]{CosmoIntertwined22}. One measure of matter clustering is the dimensionless parameter $\sigma_8$, which is the dispersion of the fractional fluctuation of the matter density $\delta = (\rho_{\rm m} - \bar{\rho}_{\rm m})/\bar{\rho}_{\rm m}$ at the present day in spheres of size $8 h^{-1}$ Mpc, if structures had grown solely by the \textit{linear} growth rate. Clustering in the late universe can be measured by weak gravitational lensing, which constrains a combination of this and the present day matter density $\Omega_{\rm m}$ as $S_8 = \sigma_8 \sqrt{\Omega_{\rm m}/0.3}$. In particular, a re-analysis of Planck satellite data \citep{Efstathiou2021} gives
\begin{equation}
\label{eq:plancks8}
S_8 = 0.828 \pm 0.016\;\; \mbox{(Planck TTTEEE)} \;\;,
\end{equation}
whereas for the late universe the Dark Energy Survey \citep[DES,][]{DES3x2} finds 
\begin{equation}
\label{eq:dess8}
S_8 = 0.779^{+0.014}_{-0.015} \;\; \mbox{(DES Y3 3x2pt)} \;\;.
\end{equation}
These are discrepant by $\sim 6\%$ ($2.3 \sigma$), and the results for other weak lensing surveys \citep{HSC3Ychi, HSC3Ycl, Kids1000} are consistent with the DES result. 
\par
One possible explanation for the difference is that the systematics of galaxy weak lensing measurements may have been underestimated: although different surveys use different analytical choices, some inherent commonality of pipeline remains between them. However, recent progress on photometric redshift calibration \citep{Hildebrandt2021, Myles2021}, shear measurements \citep{Mandelbaum2018, Kannawadi2019, MacCrann2022} and intrinsic alignments \citep[for example, see ][]{Paopiamsap2024} suggest systematics are not enough to account for the size of the difference, although different pipeline choices may lower the discrepancy to the $1.7\sigma$ level \citep{DESKids2023}.
\par 
Another possibility is that the growth of structure on \textit{linear scales} $k < 0.1 h$ Mpc$^{-1}$ differs from the $\Lambda$CDM expectation, perhaps due to modified gravity. Structure on linear scales in the late universe is also measured by the four-point correlation function of CMB temperature fluctuations, induced by weak lensing. Results from the Planck satellite \citep{Planck2020lens}, South Pole Telescope \citep{Pan2023} and the Atacama Cosmology Telescope (ACT) \citep{Darwish2021,Madhavacheril2024} are all consistent with the Planck TTTEEE power spectrum in a $\Lambda$CDM background, as stated in Equation \ref{eq:plancks8}. Although the CMB lensing sensitivity peaks between $1 < z < 3$, this range is sufficiently close to the redshift range of galaxy surveys to disfavour non-$\Lambda$CDM linear growth as an explanation.
\par
A remaining possibility is that structure on \textit{non-linear scales} $k > 0.1 h$ Mpc$^{-1}$ is different from theoretical expectations. While galaxy weak lensing pipelines discard data at highly non-linear scales (the choice of cut varies from survey to survey), they retain data for some distance into the non-linear regime. While the growth of CDM-only matter fluctuations under standard gravity is well-understood, even down to very small scales \citep{Liu2024}, other physical processes may influence the power spectrum on these scales.
\par 
In particular, the presence of baryons modifies the CDM-only expectation. The influence of baryons is qualitatively understood as 1) the suppression of structure on \textit{intermediate scales} $0.1 < k < 10^2 h$ Mpc$^{-1}$ by the pressure of \say{sub-grid} (meaning below the resolution of the hydrodynamic simulations) energetic outflows from active galactic nucleii and supernovae, and 2) the enhancement of structure on \textit{small scales} $k > 10^2 h$ Mpc$^{-1}$ by condensation due to cooling. The strength of these two effects, collectively termed baryon feedback, differs between simulations (see next Section), and the methodology to make quantitative predictions of the changes they induce in the matter power spectrum remains an active subject of research \citep[for example, see][]{hmcode2020, Eifler2015, Mohammed2018, Schneider2015JCAP, Schneider2019, Arico2021, Lu2021, Lu2022}. 
\par
In this work, we make no assumptions baryonic feedback is necessarily the only physics at work on the small-scale power spectrum. The nature of dark matter itself remains elusive. It may be \say{warm} rather than cold, may interact with itself, or may be made of ultra-light particles whose de Broglie wavelength is so large that quantum effects influence their clustering properties. These non-standard dark matter models usually suppress power on small-scales. Furthermore, it is well-known that non-zero neutrino mass suppresses small-scale power by reducing CDM-only fluctuations below the scale of their free-streaming length. These effects are difficult to disentangle from baryon feedback at the scales used in current weak lensing surveys \citep[but see][for future prospects]{Lesgourgues2024, Archidiacono2025}. 
\par
Therefore, in this paper we use a parameter that is physics-agnostic but encapsulates observational differences with the CDM-only non-linear power spectrum. \cite{Amon2022} have proposed a simple empirical model for the power spectrum as
\begin{equation}
P(k,z) = P_{\rm L} (k,z) + A_{\rm mod} \left[ P_{\rm NL} (k,z) - P_{\rm L} (k,z) \right] \;\;.
\label{eq:amon_amod}
\end{equation}
Here $P_{\rm L}$ is that predicted by the spectrum of perturbations of the CMB evolved according to linear theory in a Flat $\Lambda$CDM background expansion, and $P_{\rm NL}$ is the dark matter only non-linear power spectrum enhanced by collapsed and virialised haloes. $A_{\rm mod, I}$ is a scalar that captures suppression (or enhancement) of structure formation compared to a CDM-only benchmark. It is expressed above as being independent of scale and redshift.
\par 
Using data from DES, priors on $S_8$ and $\Omega_{\rm m}$ from Planck, and omitting the scale cuts used in the canonical analysis of DES data (see Eqn. \ref{eq:dess8}), \citet{Preston2023} find 
\begin{equation}
\label{eq:preston_result}
A_{\rm mod, I} =  0.858 \pm 0.052 \;, \; S_8 = 0.811 \pm 0.01 \;\;,
\end{equation}
where we have used the subscript I to denote this result is derived from intermediate scales.
\par 
In isolation, this result represents a repack of the $S_8$ tension by the introduction of the $A_{\rm mod}$ nuisance parameter: that the value of $S_8$ is consistent with the CMB is unsurprising given the prior. It was also not possible to say the $A_{\rm mod}$ model was preferred by the data: the $\chi^2$ fit is slightly worse compared to the same data analysed without the Planck prior and $A_{\rm mod} = 1$ \citep[see rows 4 and 6 of Table 2 of][]{Preston2023}. However, the value of $A_{\rm mod, I}$ seems consistent with the median of a range of hydrodynamical simulations \citep[see Figure 2 of][]{Preston2023} for the scales the data applies to. Similar conclusions were previously reached with a reanalysis of data from the Kilo Degree Survey (KiDs) by \citet{Amon2022}.
\par 
Our motivation in this paper is to examine the $A_{\rm mod }$ model using the weak lensing of SNe Ia. It has been recently detected at $\sim 6 \sigma$ significance that SNe Ia are weakly lensed by foreground matter \citep{Shah2024a}. The SNe Ia are dimmer if seen through voids, and brighter if seen on over-dense lines of sight (LOS). We will show in the next section that the size of this variation depends on the amplitude of the matter power spectrum on \textit{non-linear scales} rather than the linear scales described by $\sigma_8$ or $S_8$. 
\par
Weak lensing generates a non-Gaussian distribution of SN Ia magnitudes that is both generic in shape (that is, not strongly dependent on the modelling choices) and increasing in influence with redshift. As such, it may be distinguished from intrinsic non-Gaussian properties of SNe Ia (which are assumed not to vary with redshift). We forward model the probability density function (pdf) of lensing $p_{\rm lens}(\Delta m)$ conditioned on the primordial power spectrum amplitude $A_s$, the matter density $\Omega_{\rm m}$ (which is also constrained by the SN Ia Hubble diagram) and $A_{\rm mod}$. This pdf is then convolved with the intrinsic and observational noise of SN Ia data to generate a theoretical distribution of Hubble diagram residuals that may be compared to the data. We will impose priors from the CMB for $A_s$ and $\Omega_{\rm m}$ and marginalise over them, fixing the spectral index $n_s = 0.9665$ \citep{Planck2018} and the sum of neutrino masses $\Sigma m_\nu = 0.06$ eV (we discuss the sensitivity of our results to priors in Appendix \ref{sec:prior_validation}). Our result is a posterior pdf for $A_{\rm mod}$.
\par 
Our paper is organised as follows. In Section \ref{sec:theory} we describe the construction of our model. In Section \ref{sec:data} we briefly describe the data, and in Section \ref{sec:results}, we present the results of our analysis, and discuss them in Section \ref{sec:conclusions}.
\par
By linear scales, we mean $k< 0.1 h$ Mpc$^{-1}$, and we distinguish non-linear scales as intermediate: $0.1 < k< 10^2 h$ Mpc$^{-1}$, and small: $k > 10^2 h$ Mpc$^{-1}$. The reduced Hubble constant is $h = H_0 / 100$ km sec$^{-1}$ Mpc$^{-1}$. We set $c=1$ everywhere.

\section{Theory}

\label{sec:theory}

\subsection{What does the weak lensing of SNe Ia constrain?}

The parameter $\sigma_8$ is an integral measure of the amplitude of the linear power spectrum\footnote{The historical definition of $\sigma_8$ as the dispersion of density fluctuations in spheres of $8 h^{-1}$ Mpc - as proxied by galaxy counts - was motivated by the earliest galaxy surveys. It was soon understood that this was difficult to express theoretically:  fluctuations on this scale are enhanced by non-linear gravitational evolution and differ by the type of galaxy being counted due to galaxy bias. By re-defining $\sigma_8$ in terms of the \textit{linear} power spectrum, the modern definition effectively refers to structure on scales larger than $8 h^{-1}$ Mpc.}, and therefore directly relatable to the amplitude $A_s$ and shape of the primordial spectrum. 
\par 
We express the change in magnitude $\Delta m$ due to lensing of a given SN Ia as relative to a homogeneous universe of the same average matter density. $\Delta m$ can then be positive (the LOS passes through an over-dense region) or negative (through a void). To linear order, averaging over sources we have $\langle \Delta m \rangle = 0$ as gravitational lensing conserves photons. The square of the dispersion $\sigma_{\Delta m}$ of $\Delta m$ over many sources located at comoving distance $\chi_s$ may be written as an integral over the power spectrum \citep{Frieman1996}:
\begin{equation}
\label{eq:frieman}
\begin{split}
\sigma_{\Delta m}^2 = & \;\; 9\pi(0.4\log{10})^{2} \; \Omega_{\rm m}^{2} H_0^4 \;\times \\
& \int_{0}^{\chi_s} d\chi \frac{\chi^2 (\chi_s-\chi)^2}{\chi_s^2} (1+z(\chi))^2 
 \int_{0}^{k_{\rm max}} dk \frac{\Delta^2(k,z)}{k^2}  \;\;,
\end{split}
\end{equation}
where $\Delta^2(k,z) =k^3 P(k,z)/2\pi^2$ is the dimensionless matter power spectrum, and we have chosen to cut off the scale at $k_{\rm max}$. The pre-factors of $(0.4\log{10})$ arise as $\sigma_{\Delta m}$ is the dispersion in magnitudes rather than flux amplification. The value of this integral depends considerably on the power spectrum at small scales.
\par
\begin{figure}
    \centering
    \includegraphics[width=\columnwidth]{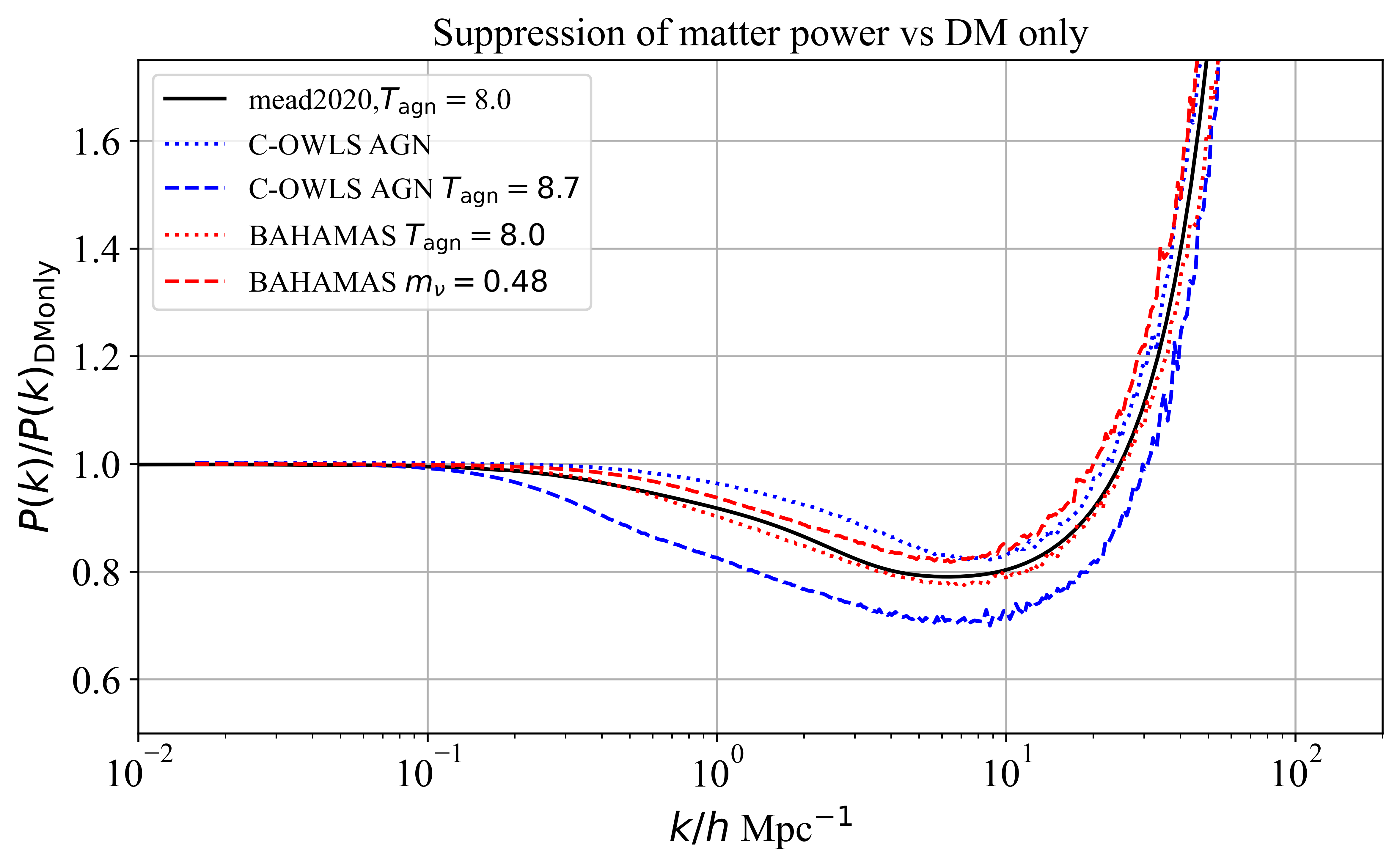} 
    \caption{The suppression or enhancement of the power spectrum compared to the dark matter only model of \citet{Mead2020}. The models used are \texttt{HMCODE2020} with $T_{\rm AGN} = 8.0$, the Cosmic-OWLS hydrodynamical simulations \citep{LeBrun2014}, and the BAHAMAS hydrodynamical simulations \citep{McCarthy2017}. Power is suppressed on scales $0.1 < k/h < 30 $ Mpc$^{-1}$ by AGN and supernovae feedback prescriptions, which differ from model to model. At scales  $k/h > 30 $ Mpc$^{-1}$ power is enhanced due to condensation from baryonic cooling, although again the extent of this depends considerably on the model.}
    \label{fig:suppression}
\end{figure}

 
To illustrate this for the case of baryon feedback, we select a range of models from the \texttt{HMCODE2020}\footnote{\url{https://github.com/alexander-mead/HMcode}} emulation package\footnote{As implemented in \texttt{CAMB}, \url{https://github.com/cmbant/CAMB}}. We choose three options: 1) linear only, 2) the CDM-only non-linear model of \citet{Mead2020}, and 3) the baryon feedback model of the same with $T_{\rm AGN} = 8.0$. While \texttt{HMCODE2020} is not designed to extrapolate to the $k_{\rm max}$ relevant for us, it nevertheless provides a useful illustration of uncertainty on small scales. To this, we add power spectrum data\footnote{\url{https://powerlib.strw.leidenuniv.nl}} compiled from hydrodynamical simulations by \citet{VanDaalen2020}. The simulations we select are the Cosmic-OWLS suite \citep{LeBrun2014} and the BAHAMAS suite \citep{McCarthy2017}, which encompass a reasonable range of outcomes. We plot the ratio of the baryon feedback power spectra at $z=0$ to the relevant CDM-only reference model in Figure \ref{fig:suppression}. The figure shows power is suppressed up to scales of $k \sim 30 h$ Mpc $^{-1}$, then strongly enhanced with considerable difference between models. Equivalent figures for non-standard dark matter models and neutrino masses can be found in \citet{Lesgourgues2024, Archidiacono2025}.
\par
\begin{figure}
    \centering
    \includegraphics[width=\columnwidth]{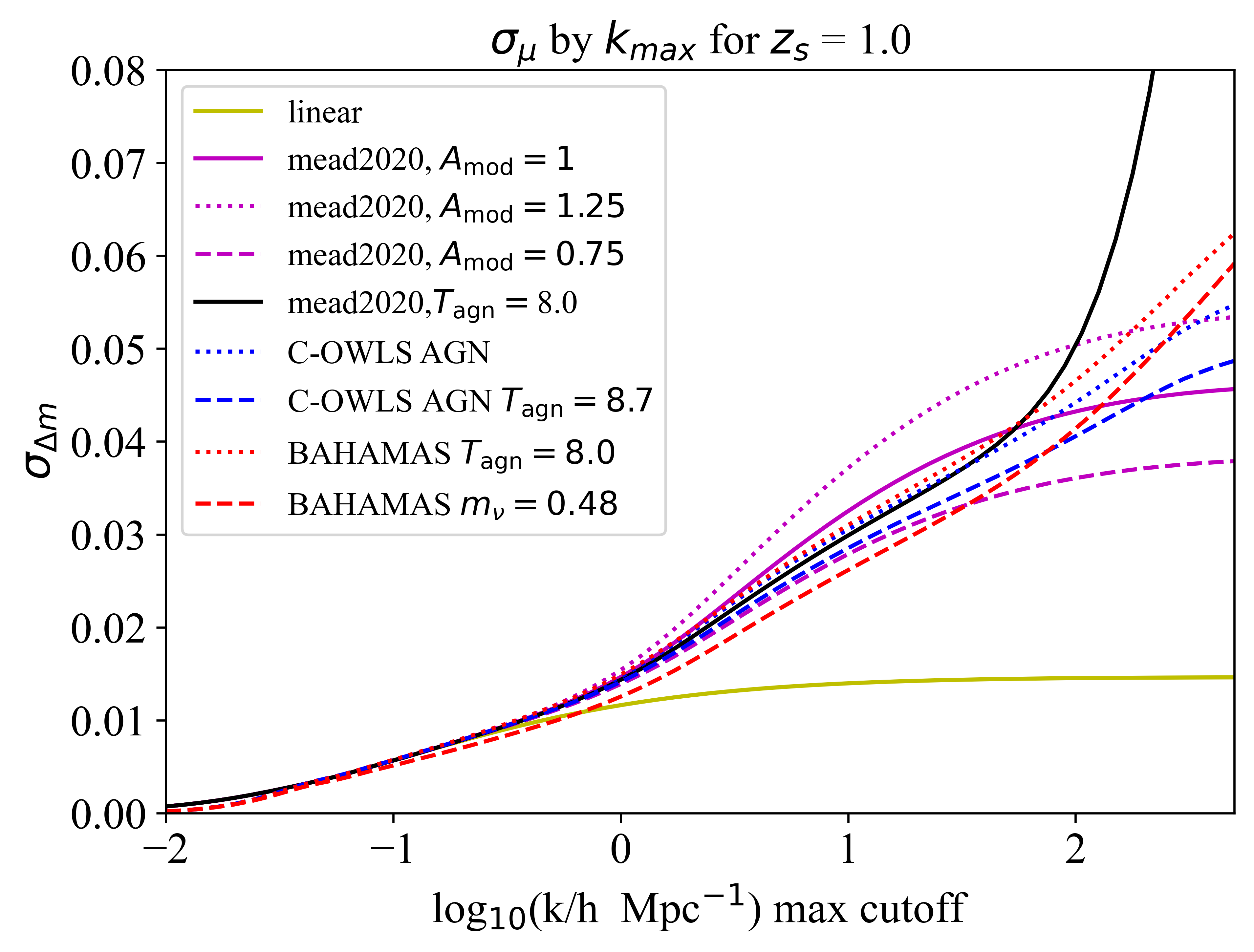} 
    \caption{The dispersion of lensing magnification derived from Eqn. \ref{eq:frieman} as a function of the integral cutoff $k_{\rm max}$ with the lens at $z_l =0.5$ and the source at $z_s = 1.0$, for a range of models and simulations of the matter power spectrum. The linear power spectrum is shown in yellow, the model of \citet{Mead2020} is shown without baryon feedback in magenta and with feedback in black Dotted and dashed magenta lines show the model adjusted for two choices of $A_{\rm mod}$, as defined in Equation \ref{eq:amon_amod}. Results from power spectra compiled from hydrodynamical simulations and published in \citet{VanDaalen2020}, for a selction of parameters, are shown in red and blue. It is clear that the value of $\sigma_{\Delta m}$ is sensitive to the power spectrum on intermediate and small scales. In particular, while baryonic feedback suppresses lensing on intermediate scales, it enhances it on the small scales relevant to SN Ia. It is not clear that all of the models converge at small scales: in the case of \citet{Mead2020} we have extrapolated the emulator far beyond the region it was designed to model. In the case of simulation data, it is likely that the softening length and particle size produce spurious additional power on small scales. This graph is for illustration only and does not form part of our analysis.}
    \label{fig:sigm_by_kmax}
\end{figure}

In Figure \ref{fig:sigm_by_kmax} we plot $\sigma_{\Delta m}$ as a function of the small scale cutoff $k_{\rm max}$ for the same selection of models. There are considerable differences between models: while the dark matter only model has little sensitivity to scales $k > 10^2 h$ Mpc $^{-1}$, results from baryonic feedback or neutrino prescriptions vary considerably. The apparent runaway behaviour of the model of \cite{Mead2020} is due to a flattening of the power spectrum at $k > 10^2 h$ Mpc $^{-1}$ because the stellar component is represented as a delta function (the authors make no claims their model can be extrapolated beyond $k \ sim 20$). The simulation results start to diverge at $k > 250 h$ Mpc $^{-1}$. This is likely due to the softening lengths ($r = 4$ kpc for Cosmic-OWLS and BAHAMAS) and particle masses ($4 \times 10^9$ - $7 \times 10^8 M_{\odot}$) used in them which lead to spurious clumping on small scales. We estimate that to obtain adequate predictions of SN Ia lensing from N-body or hydrodynamical simulations would require particle masses of $m_p \sim 10^7 M_{\odot}$ and softening lengths of 1 kpc (corresponding to $k \sim 10^3 h $ Mpc$^{-1}$). This applies either to estimation from the power spectrum or directly through ray-tracing. We conclude that a reliable a-priori calculation of $\sigma_{\Delta m}$ is at present unavailable, and we emphasize that we do not make use of any of these models or Eqn. \ref{eq:frieman} in our analysis. 
\par 
It is also clear from Fig. \ref{fig:sigm_by_kmax} that SN Ia weak lensing is insensitive to the linear regime and associated cosmological parameters such as $\sigma_8$. Indeed, linear-scale correlation is undetectable with current datasets \citep{Shah2024a}. In this paper, we will fix the linear scales using a prior on $A_s$ from the CMB, and use SNe Ia to constrain the non-linear empirical amplitude $A_{\rm mod}$.
\par
Before continuing, we note that the existence of compact objects (CO, for example primordial black holes, if they exist) produces an enhancement in $\sigma_{\Delta m}$ if they form a significant fraction $\alpha = \Omega_{\rm CO} / \Omega_{\rm m}$ of the matter density. In \citet{Shah2024b}, it was shown that $\alpha < 0.12 $ at 95\% credibility, and $\alpha = 0$ was preferred (by the Bayes ratio of model probabilities) to $\alpha > 0$. Hence, while some compact objects (e.g. stars) certainly exist, they do not make a measurable contribution to SN Ia lensing. We therefore neglect them in this analysis.
\par

\subsection{SN Ia weak lensing as a function of $A_{\rm mod}$}
In the halo model description, matter power is the sum of linear low density-contrast perturbations, plus gravitationally bound high density-contrast haloes that have collapsed and virialised \citep{Kaiser1984}. This is also taken as the starting point for the codes used in weak lensing shear analyses such as \texttt{HMCODE2020}, which adjusts the theoretical halo-model power spectrum empirically to better fit simulations \citep{Mead2020}. We write our power spectrum model as denoted by subscripts L and H respectively: 
\begin{equation}
P(k,z) = P_{\rm L} (k,z) + A_{\rm mod} P_{\rm H} (k,z) \;\;.
\label{eq:my_amod}
\end{equation}
$P_{\rm L}$ is the same as that in Eqn. \ref{eq:amon_amod}, while $P_{\rm H}$ is the contribution solely due to halos, and equivalent to the term $P_{\rm NL} - P_{\rm L}$ in Eqn. \ref{eq:amon_amod}. For $P_{\rm H}$ we adopt the calibration of \citet{Sheth2001} refitted by \citet{Courtin2011}, which describes the abundance and spectrum of haloes in a purely CDM universe very well over a huge range of halo masses \citep{Zheng2024}. The calibration of \citet{Sheth2001} was also used as the starting point for power spectrum emulator models such as \citet{Smith2003} and successors. We take the haloes to have the NFW profile described in \citet{Navarro1997}, as it was shown in \citet{Shah2024a} that the NFW model is consistent with observations of SN Ia lensing. Although in general the profiles will not be spherically symmetric, this is an accurate approximation after averaging over many LOS \citep{Mandelbaum2005}. 
\par
Our aim is to model the statistics of magnitude fluctuations of SNe Ia in terms of $A_{\rm mod}$. Before we proceed to the details below, we note that whilst our $A_{\rm mod}$ is equivalent in formulation to that of \citet{Amon2022}, important practical differences arise when applied to SNe Ia data. The integral of Eqn. \ref{eq:frieman} does not define a window function of scale other than limits imposed by the size of the source and lens, and other statistics of SN Ia lensing operate in the same fashion. This is distinct from the scales introduced by spatial correlations functions of transverse separation used in galaxy surveys, which may then be used to set scale cuts. Fig. \ref{fig:sigm_by_kmax} indicates that for intermediate scales, there is a broad expectation that baryonic feedback will result in $A_{\rm mod} < 1$ \citep[the models considered generate less lensing than the CDM-only model of][]{Mead2020}. Conversely, continuing to small scales relevant for SN Ia lensing, it is likely that baryons alone would lead to $A_{\rm mod} > 1$ (more lensing is observed than expected in a CDM-only universe), although we noted above that the resolution limits of the simulations may generate spurious power enhancement on these scales. Alternatively, warm dark matter (if it exists) and neutrino masses are expected to suppress power on small scales, reducing $A_{\rm mod}$. We therefore have no a-priori expectations of whether the $A_{\rm mod, S}$ we measure from SNe Ia will indicate power suppression or enhancement. 
\par
Following the procedure of \citet{Zumalacarregui2018}, revised in \citet{Shah2024b}, we model the full-shape of the lensing pdf\footnote{We avoid using methods based on higher order moments of the pdf \citep[such as described in][]{Marra2013,Quartin2014}, which are vulnerable to the presence of outliers.} as a function of cosmological parameters, intrinsic skew and $A_{\rm mod}$. We write the lensing pdf as a convolution of lensing due to linear scales and haloes as
\begin{equation}
p_{\rm lens}(\Delta m) = p_{\rm L} (\Delta m ; A_s, \Omega_{\rm m}, z) * p_{\rm H} (\Delta m ; A_s, \Omega_{\rm m}, A_{\rm mod}, z) \;\;,
\end{equation}
where $*$ denotes the convolution operation. The 1-point distribution of weak lensing convergence on linear scales has been shown to be well-approximated by a log-normal distribution \citep{Clerkin2017}. We therefore take $p_{\rm L}$ as a log-normal distribution of zero mean and dispersion $\sigma_{\rm L}$ obtained from Eqn. (\ref{eq:frieman}) using the linear power spectrum. We obtain $p_{\rm H}$ from \texttt{TurboGL}\footnote{\url{https://github.com/valerio-marra/turboGL}} \citep{Kainulainen2009, Kainulainen2011a} which uses semi-analytic integration to accurately model lensing by dark matter haloes (and thus avoids the resolution issues inherent in N-body or hydrodynamical simulations). The minimum halo mass has been set to be $10^7 M_{\odot}$. $A_{\rm mod}$ is then a simple scale parameter on this pdf such that $\sigma_{\rm H}(A_{\rm mod}) = A_{\rm mod} \sigma_{\rm H}(A_{\rm mod}=1)$.
\par
Intrinsic skew of SN Ia residuals may in principle be confused with lensing. However, while lensing skew is redshift-dependent, intrinsic skew is presumed not to be. We parametrise the intrinsic dispersion of SN Ia by the sin-arcsin distribution family \citep{Jones2009} where $\delta, \epsilon$ of this family capture both skew and kurtosis with $\delta=1, \epsilon =0$ being a normal distribution. The distribution is defined as
\begin{equation}
p_{\rm Int}(\Delta m ) = \frac{1}{\sqrt{2\pi} E} \delta \sqrt{1+x^2} \exp(-x^2/2)/\sqrt{1+\Delta m^2} \;\;,
\end{equation}
where 
\begin{equation}
x  = (\sinh{(\delta \mbox{arcsinh}{(\Delta m)} - \epsilon)} - D)/E \;\;,
\end{equation}
and the location and scale parameters $D,E$ are determined by the constraints
\begin{equation}
\label{eq:sigmai}
\begin{split}
\int \Delta m p_{\rm Int}(\Delta m) d\Delta m  & = 0 \\
\int \Delta m^2 p_{\rm Int}(\Delta m) d\Delta m  & = \sigma_{i}^{2} \;\;.
\end{split}
\end{equation}
Here $\sigma_i^2 = C_{ii}$ is the diagonal of the SN Ia covariance matrix (see Section \ref{sec:lenslike} below), which is the statistical uncertainty in the SN Ia distance modulus. Convolving this with the lensing pdf gives   
\begin{equation}
\label{eq:pres}
p_{\rm res} (\Delta m)= p_{\rm lens} * p_{\rm Int} \;\;,
\end{equation}
with $\Delta m$ now the Hubble diagram residual 
\begin{equation}
\Delta m = m_i - \mathcal{M} - \mu_{i, \rm theory}(\theta) \;\;.
\end{equation}
Here $\mathcal{M} = M - 5\log_{10} H_0$ is a degenerate combination of the Hubble constant $H_0$ and the fiducial SN Ia absolute magnitude $M$. The distance modulus is $\mu_{i, \rm theory} = 5 \log_{10} (D_{\rm L}(z_i, \Omega_{\rm M}) H_0/c ) + 25$ and we take $D_{\rm L}$ as the usual homogeneous cosmology luminosity distance in a Flat $\Lambda$CDM model.

\subsection{The SN Ia likelihood adjusted for lensing}
\label{sec:lenslike}
Denoting the model parameters collectively as $\theta$, the SN Ia likelihood for a homogeneous cosmology using the DES SN Ia Year-5 sample (DES-SN5YR) is assigned to be the Gaussian  
\begin{equation}
    \ln \mathcal{L}_{G} (\Delta m | \theta) = -\frac{1}{2} \sum_{i,j} \Delta m_{i} C^{-1}_{ij} \Delta m_{j} \;\;
\end{equation}
where $C$ is the covariance matrix which is the sum of systematic and statistical errors \citep{Vincenzi2024}. In the presence of non-Gaussian lensing, we adjust this to 
\begin{equation}
\label{eq:lenslike}
\log \mathcal{L}  = \ln \mathcal{L}_{G}  + \left( \sum_i \log p_{\rm res}- \sum_i \log p_{\rm diag}\right) \;\;,
\end{equation}
with  
\begin{equation}
p_{\rm diag}(\Delta m_i) = \frac{1}{\sqrt{2\pi} \sigma_i } \exp (-\frac{1}{2}(\Delta m_i/\sigma_i)^2) \;\;,
\end{equation}
where $\sigma_i^2 = C_{ii}$ as per Eqn. \ref{eq:sigmai} and $p_{\rm res}$ is defined in Eqn. \ref{eq:pres}. The term in brackets of Eqn. \ref{eq:lenslike} adjusts the likelihood of each individual SN Ia for the difference between a skewed (either by lensing or intrinsic skew) residual probability and a Gaussian probability. The first term retains covariance, which is important for correct error estimation. As explained in \citet{Shah2024b}, this likelihood reduces correctly when there is no lensing or intrinsic skew as in this case $p_{\rm res} = p_{\rm diag}$, and also when the covariance matrix is diagonal as in this case $\log \mathcal{L}_{G}  = \sum_i \log p_{\rm diag}$. The key point is $C$ is only weakly non-diagonal, so we expect our assignment to be accurate. We discuss the validation of this likelihood in Appendix \ref{sec:likelihood_validation}.
\par
We pre-compute the lensing pdfs for a grid of redshifts, cosmological parameters and $A_{\rm mod}$, and interpolate the log probabilities. The evaluation of the likelihood takes $\sim 0.3$ seconds for $\sim 1,500$ SN Ia on a typical laptop. We fix the optical depth of reionisation as $\tau = 0.0561$ and the power spectrum slope as $n_s = 0.9665$, and set the neutrino mass $m_\nu = 0.06$ eV. We take uniform priors as $A_{\rm mod} \in (0.2, 2.5)$ (motivated by consideration of Fig. \ref{fig:sigm_by_kmax}), $\epsilon \in (-0.2,0.2)$ and $\delta \in (0.6, 1.4)$. Runs are performed using \texttt{Polychord}\footnote{\url{https://github.com/PolyChord/PolyChordLite}} \citep{Polychord}, and plots and analysis are made using \texttt{Anesthetic}\footnote{\url{https://github.com/handley-lab/anesthetic}} \citep{Anesthetic}.

\section{Data}
\label{sec:data}
\subsection{SN Ia}

We use the DES-SN5YR dataset as described in \citet{Sanchez2024}, but with the modification not to exclude SNe Ia that are more than $4 \sigma$ away from the best-fit Hubble diagram. Removing this cut avoids biasing our results by arbitrarily truncating the skewed and extended distribution of residuals that lensing produces. The SN Ia survey was conducted in four regions of the DES footprint with a total of ten fields, and the SN Ia range from $ 0.01 < z < 1.13$.  Supernova candidates are analysed using machine-learning classifiers \citep{Moeller2020, Moeller2022, Qu2021} whose inputs are the light curve observations, and whose output is the probability of being an SN Ia. The diagonal of the covariance is then adjusted for this probability, down-weighting likely contaminants but not discarding them altogether. The SN Ia redshift is set to be the spectroscopic redshift of the galaxy that is closest in directional light radius to the SN Ia \citep{Sullivan2006, Qu2024}. 
\par
There are 1,930 SN in the initial sample, of a similar redshift distribution to the original DES-SN5YR dataset, and we cut these to include only those between  $0.2 < z< 1.0$ in our analysis. The lower cut is because lensing will not materially affect low redshift SN Ia, and the lower redshift SN Ia are from older, heterogeneous surveys with uncertain selection functions. The upper cut is to reduce potential uncertainties due to larger bias corrections at high redshifts \citep[for example, see Figure 7 of ][]{Vincenzi2024}. We additionally cut likely contaminants or poorly measured SN Ia, by excluding data with $\sigma_{m} > 1.0 $ mag and $p$(SN Ia) $< 0.9$. Our data input to the lensing likelihood  therefore comprises 1,484 SN Ia of average redshift $ z \sim 0.47$.

\subsection{CMB}
As input priors to $H_0, \Omega_{\rm m}$ and linear scales $A_s$, we use chains derived from the Python implementation of Planck’s 2015 \texttt{Plik\_lite} \citep{prince19} which may be found on the repository of the DES-SN5YR data\footnote{\url{https://github.com/des-science/DES-SN5YR}}. These chains have somewhat wider constraints than the fiducial results of \cite{Planck2018}, but are perfectly adequate for our results. 

\section{Results}
\label{sec:results}
\begin{figure*}
    \centering
    \includegraphics[width=\textwidth]{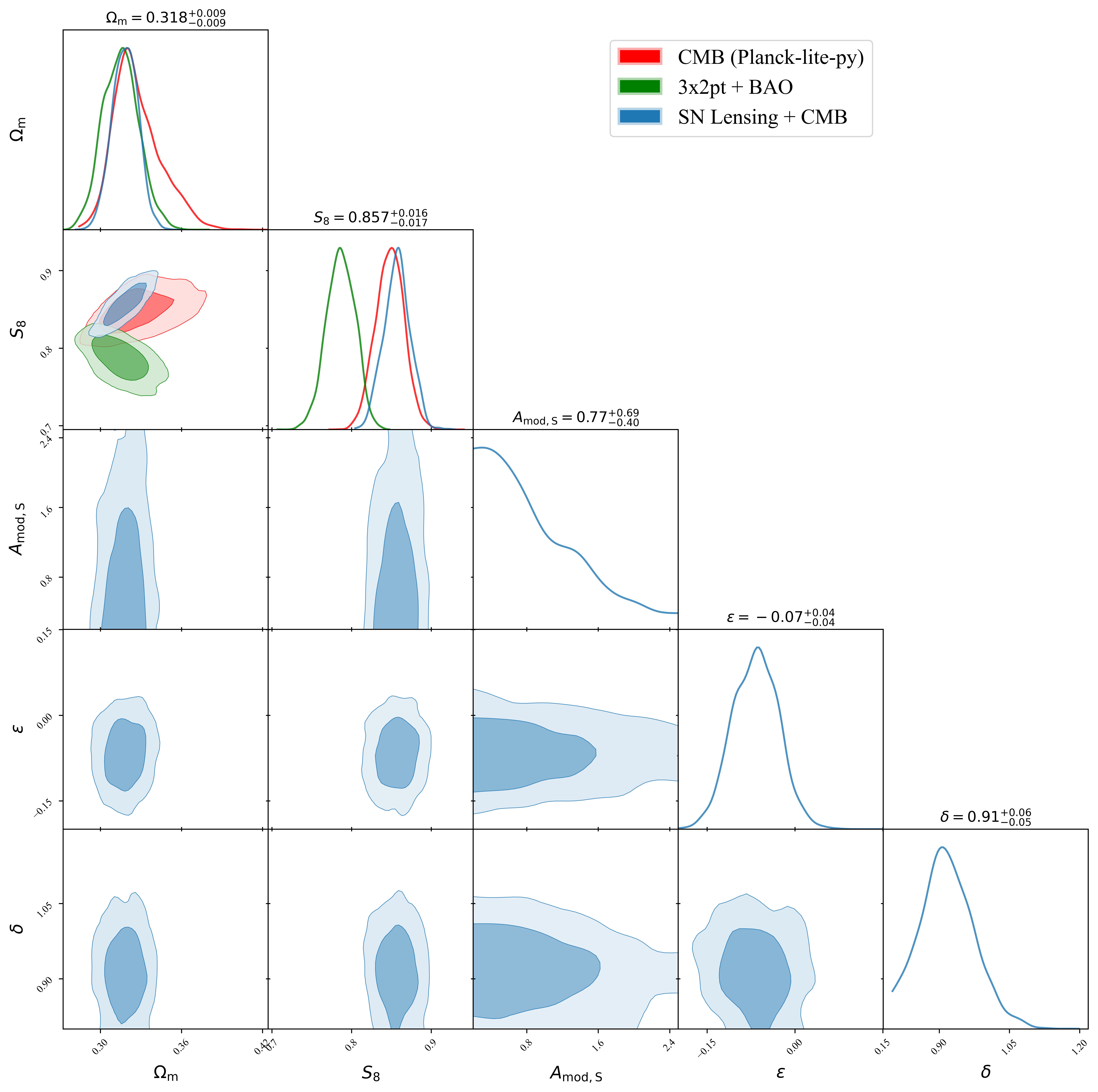}
    \caption{A triangle plot of the posteriors for relevant model parameters, with the medians and 68\% quantiles shown along the diagonal. The constraints on $S_{8}$ and $\Omega_{\rm m}$ arise from the fit of the CMB power spectrum and SN Ia luminosity distances.  The CMB priors we use are shown in red. As noted in the text, the Planck-lite-py likelihood used to generate the chains has moderately wider constraints than the full likelihood used in \citet{Planck2018}, however this does not impact our results. In contrast, the constraint on $A_{\rm mod, S}$ arises from the detail of the distribution of SN Ia residuals around the mean and is not degenerate with $S_8$. $\epsilon$ and $\delta$ represent intrinsic skew and kurtosis of the distribution; it is evident from the posterior that there is little degeneracy between this and lensing. For comparison, we also plot a combination (in green) of the DES 3x2pt analysis of \citet{DES3x2} and BAO measurements from the Sloan Digital Sky Survey (SDSS), as summarised in \citet{Alam2021}. We remind the reader that we have defined $A_{\rm mod}=1$ as small-scale power that is compatible with the growth of CDM-only fluctuations in a Flat $\Lambda$CDM universe with cosmological parameters derived from the CMB.}
    \label{fig:snbao_triangle}
\end{figure*}

Marginalising over all other parameters, and quoting the median, 16\% and $84\%$ quantiles, we find 
\begin{equation}
    A_{\rm mod, S} = 0.77^{+0.69}_{-0.40}
\end{equation} which is consistent within 1$\sigma$ of the CDM-only value of $A_{\rm mod} = 1$, and we have used the subscript S to denote our result is derived primarily from small scales. We show the triangle plot for the marginalised posterior and pair distributions in Figure \ref{fig:snbao_triangle}.
\par 
There are hints the data prefers lower values with the maximum of the full posterior (as determined by a kernel density estimate) at $A_{\rm mod, S} = 0.30$, and the highest-density 68\% credible interval is $A_{\rm mod, S} < 1.09$. We expect this upper bound to be conservative, as Figure \ref{fig:snbao_triangle} shows our posterior is truncated by our prior and it is likely that more probability mass might be found below our prior range rather than above it. Conversely, our credibile interval around the median will be somewhat over-optimistic for the same reason.
\par
Notably, our method demonstrates that SN Ia can distinguish the effects of the linear and non-linear parts of the power spectrum, as $S_8$ and $A_{\rm mod, S}$ show little correlation in their posteriors. The credible intervals for the intrinsic skew parameters $\epsilon = -0.07^{+0.04}_{-0.04}$ and $\delta = 0.91^{+0.06}_{-0.05}$ are moderately discrepant from the values $(0,1)$ of a Gaussian distribution, but at no great significance. There is little covariance between intrinsic skew and matter-power parameters. 
\par
Our maximum likelihood improves the fit over the Gaussian likelihood by $\delta \chi^2 = -5.6$. Subtracting out the fit improvement due to the introduction of intrinsic skew, we find $A_{\rm mod, S} = 0.3$ improves the fit by $\delta \chi^2 = -1.2$ compared to $A_{\rm mod, S} = 1$.
\par 
We describe our estimation of systematics in Appendix \ref{sec:validation}, with the dominant contribution being the assignment of the likelihood. While there is some inaccuracy in our posterior, examining the expected coverage probability we see that the 68\% highest-density credible interval we quote above is likely to be conservative (over and above the effect of truncation by the prior noted above) by $\delta A_{\rm mod, S} \sim 0.19$. Other credible intervals of interest have systematics of a similar amount or smaller. We have tested the effect of the redshift and SN Ia probability cuts used for our data and find they are small. Our total systematic error estimate is $\delta A_{\rm mod, S} \sim 0.2$. Therefore, we judge that systematics are smaller than our statistical error by a factor of at least two.

\par
The literature has to date quoted results in terms of $\sigma_8$ which (ignoring for now the caveats we noted in Section \ref{sec:theory}) this be interpreted as $\sim A_{\rm mod, S} \sigma_{8,\rm CMB}$. The first measurement was done by \citet{Castro2014}, who found $\sigma_8 = 0.84^{+0.28}_{-0.65}$ using 706 SN Ia from the Joint Lightcurve (JLA) catalogue \citep{Betoule2014}. Re-analysing the JLA data using a model that also incorporated SN Ia peculiar velocities (via their redshifts), \citet{MacAulay2017} found similar results but also that the systematics in \citet{Castro2014} were underestimated. In \citet{MacAulay2020}, 196 SN Ia from the DES Year 3 release were used in the same methodology to find $\sigma_8  = 1.2^{+0.9}_{-0.8}$. Calibrating a halo model using the observed correlation between SN Ia residuals and foreground galaxy positions, \citet{Shah2024a} found $\sigma_8  = 0.9 \pm 0.13$ by comparing the dispersion $\sigma_{\Delta m}$ of lensing in this model along random lines of sight to a fitting formula given in \cite{Marra2013}. In general, although these results do not marginalise over cosmological parameters and intrinsic skew as we have done, they all indicate a weak preference for $A_{\rm mod, S} > 1$. However, they are consistent with our results within $1 \sigma$.

\section{Conclusions}
\label{sec:conclusions}
In this paper, we have shown that the non-Gaussian distribution of residuals of SNe Ia to the Hubble diagram induced by weak gravitational lensing carries statistical information about the matter power spectrum on scales $k>1 h$ Mpc$^{-1}$. In particular, SNe Ia data provides access to scales that cannot be probed by galaxy-sized sources. Due to the theoretical uncertainties of modelling the power spectrum on these scales, we have constrained an empirical parameter $A_{\rm mod, S}$, which describes the suppression (or enhancement) of matter power on small scales compared to a benchmark of a dark matter only $\Lambda$CDM universe with cosmological parameters derived from the CMB. We find hints of suppression with $A_{\rm mod, S} = 0.77^{+0.69}_{-0.40}$ (median and 68\% credible interval), with the posterior peaking at low values of $A_{\rm mod, S}$. 
\par 
While our results are weaker than those of \citet{Preston2023} (quoted as Eqn. \ref{eq:preston_result} here), who find almost $3\sigma$ preference for $A_{\rm mod} <1 $, they are independent of data and the typical systematics of galaxy shear surveys such as photometric redshifts, shape blending and intrinsic alignments. By contrast, our systematics arise primarily from assignment of the likelihood. 
\par 
Our results are not precise enough to distinguish between competing models of dark matter or baryonic physics. The pathway to doing so lies in improvements to statistical error, systematics and theoretical modelling. Regarding statistical error, in \citet{Quartin2014}, it was forecast that a sample of 3,000 SN Ia from DES would be able to constrain $\sigma_8$ to within $\sim 35\%$; this is consistent with our $\sim 55\%$ constraint for $A_{\rm mod}$ using 1,484 SN Ia. The authors also forecast that a sample of 500,000 SN Ia from the Rubin LSST survey would result in a constraint of $\sim 3\%$. In this case, the statistical error would be below the systematics, and the current likelihood assignment would need to be improved. One way this could be achieved is the use of simulation-based inference. 
\par
Finally, regarding theoretical modelling it would be desirable to interpret our results in the context of parameters of physical processes, which requires extending these models to small scales. As noted in Section \ref{sec:theory}, the apparent non-convergence of the baryonic models illustrated in Fig. \ref{fig:sigm_by_kmax} is likely due to an extrapolation of emulators beyond the scales they were trained on, because of the resolution limits of hydrodynamical simulations. This is in principle surmountable: promising work has been done on \say{nesting} dark matter only simulations to increase resolution \citep{Wang2020, Zheng2024} arbitrarily, and it seems plausible that this methodology could be applied to other simulations. Even if this were not possible, we note that certain ranges of $A_{\rm mod}$ would provide severe constraints on the underlying physics: for example, $A_{\rm mod} \lesssim 0.5$ is not expected from any of the models considered here with cold dark matter as the dominant component. 
\par 
We conclude that for future datasets, SN Ia offer a unique window into the power spectrum on small scales, and the pathway to improve the control of systematics and theoretical modelling is clear.

\section*{Acknowledgements}

P.S. devised the project, performed the analysis and drafted the manuscript; M.V. re-analysed the DES-SN5YR data with cuts removed; T.M.D., R.C., D.H., L.G., N.J., J.L., C.L., M.Su., L.W. advised on the analysis and commented on the manuscript; P.A. assisted with the Pippin pipeline used for the simulations. T.M.D. and M.Su. were also internal reviewers and R.M. was the final reader. The remaining authors have made contributions to this paper that include, but are not limited to, the construction of DECam and other aspects of collecting the data; data processing and calibration; developing broadly used methods, codes, and simulations; running the pipelines and validation tests; and promoting the science analysis.

This paper has gone through internal review by the DES collaboration, and we thank the anonymous referee for his helpful comments. 

{\footnotesize 
We acknowledge the following former collaborators, who have contributed directly to this work --- Ricard Casas, Pete Challis, Michael Childress, Ricardo Covarrubias, Chris D'Andrea, Alex Filippenko, David Finley, John Fisher, Francisco Förster, Daniel Goldstein, Santiago González-Gaitán, Ravi Gupta, Mario Hamuy, Steve Kuhlmann, James Lasker, Marisa March, John Marriner, Eric Morganson, Jennifer Mosher, Elizabeth Swann, Rollin Thomas, and Rachel Wolf.

T.M.D., A.C., R.C., acknowledge the support of an Australian Research Council Australian Laureate Fellowship (FL180100168) funded by the Australian Government, and A.M. is supported by the ARC Discovery Early Career Researcher Award (DECRA) project number DE230100055.
M.S. and J.L are supported by DOE grant DE-FOA-0002424 and NSF grant AST-2108094.
R.K.\ is supported by DOE grant DE-SC0009924. M.V.\ was partly supported by NASA through the NASA Hubble Fellowship grant HST-HF2-51546.001-A awarded by the Space Telescope Science Institute, which is operated by the Association of Universities for Research in Astronomy, Incorporated, under NASA contract NAS5-26555. 
L.G. acknowledges financial support from AGAUR, CSIC, MCIN and AEI 10.13039/501100011033 under projects PID2023-151307NB-I00, PIE 20215AT016, CEX2020-001058-M, and 2021-SGR-01270. D.S. was supported in part by NASA grant 14-WPS14-0048. The UCSC team is supported in part by NASA grants NNG16PJ34G and NNG17PX03C issued through the Roman Science Investigation Teams Program; NSF grants AST-1518052 and AST-1815935; NASA through grant No. AR-14296 from the Space Telescope Science Institute, which is operated by AURA, Inc., under NASA contract NAS 5-26555; the Gordon and Betty Moore Foundation; the Heising-Simons Foundation.
We acknowledge the University of Chicago’s Research Computing Center for their support of this work. 

Funding for the DES Projects has been provided by the U.S. Department of Energy, the U.S. National Science Foundation, the Ministry of Science and Education of Spain, 
the Science and Technology Facilities Council of the United Kingdom, the Higher Education Funding Council for England, the National Center for Supercomputing 
Applications at the University of Illinois at Urbana-Champaign, the Kavli Institute of Cosmological Physics at the University of Chicago, 
the Center for Cosmology and Astro-Particle Physics at the Ohio State University,
the Mitchell Institute for Fundamental Physics and Astronomy at Texas A\&M University, Financiadora de Estudos e Projetos, 
Funda{\c c}{\~a}o Carlos Chagas Filho de Amparo {\`a} Pesquisa do Estado do Rio de Janeiro, Conselho Nacional de Desenvolvimento Cient{\'i}fico e Tecnol{\'o}gico and 
the Minist{\'e}rio da Ci{\^e}ncia, Tecnologia e Inova{\c c}{\~a}o, the Deutsche Forschungsgemeinschaft and the Collaborating Institutions in the Dark Energy Survey.

The Collaborating Institutions are Argonne National Laboratory, the University of California at Santa Cruz, the University of Cambridge, Centro de Investigaciones Energ{\'e}ticas, 
Medioambientales y Tecnol{\'o}gicas-Madrid, the University of Chicago, University College London, the DES-Brazil Consortium, the University of Edinburgh, 
the Eidgen{\"o}ssische Technische Hochschule (ETH) Z{\"u}rich, 
Fermi National Accelerator Laboratory, the University of Illinois at Urbana-Champaign, the Institut de Ci{\`e}ncies de l'Espai (IEEC/CSIC), 
the Institut de F{\'i}sica d'Altes Energies, Lawrence Berkeley National Laboratory, the Ludwig-Maximilians Universit{\"a}t M{\"u}nchen and the associated Excellence Cluster Universe, 
the University of Michigan, NSF's NOIRLab, the University of Nottingham, The Ohio State University, the University of Pennsylvania, the University of Portsmouth, 
SLAC National Accelerator Laboratory, Stanford University, the University of Sussex, Texas A\&M University, and the OzDES Membership Consortium.

Based in part on observations at Cerro Tololo Inter-American Observatory at NSF's NOIRLab (NOIRLab Prop. ID 2012B-0001; PI: J. Frieman), which is managed by the Association of Universities for Research in Astronomy (AURA) under a cooperative agreement with the National Science Foundation.
 Based in part on data acquired at the Anglo-Australian Telescope. We acknowledge the traditional custodians of the land on which the AAT stands, the Gamilaraay people, and pay our respects to elders past and present. Parts of this research were supported by the Australian Research Council, through project numbers CE110001020, FL180100168 and DE230100055. Based in part on observations obtained at the international Gemini Observatory, a program of NSF’s NOIRLab, which is managed by the Association of Universities for Research in Astronomy (AURA) under a cooperative agreement with the National Science Foundation on behalf of the Gemini Observatory partnership: the National Science Foundation (United States), National Research Council (Canada), Agencia Nacional de Investigaci\'{o}n y Desarrollo (Chile), Ministerio de Ciencia, Tecnolog\'{i}a e Innovaci\'{o}n (Argentina), Minist\'{e}rio da Ci\^{e}ncia, Tecnologia, Inova\c{c}\~{o}es e Comunica\c{c}\~{o}es (Brazil), and Korea Astronomy and Space Science Institute (Republic of Korea).  This includes data from programs (GN-2015B-Q-10, GN-2016B-LP-10, GN-2017B-LP-10, GS-2013B-Q-45, GS-2015B-Q-7, GS-2016B-LP-10, GS-2016B-Q-41, and GS-2017B-LP-10; PI Foley).  Some of the data presented herein were obtained at Keck Observatory, which is a private 501(c)3 non-profit organization operated as a scientific partnership among the California Institute of Technology, the University of California, and the National Aeronautics and Space Administration (PIs Foley, Kirshner, and Nugent). The Observatory was made possible by the generous financial support of the W.~M.~Keck Foundation.  This paper includes results based on data gathered with the 6.5 meter Magellan Telescopes located at Las Campanas Observatory, Chile (PI Foley), and the Southern African Large Telescope (SALT) (PIs M.~Smith \& E.~Kasai).
The authors wish to recognize and acknowledge the very significant cultural role and reverence that the summit of Maunakea has always had within the Native Hawaiian community. We are most fortunate to have the opportunity to conduct observations from this mountain.

The DES data management system is supported by the National Science Foundation under Grant Numbers AST-1138766 and AST-1536171.
The DES participants from Spanish institutions are partially supported by MICINN under grants ESP2017-89838, PGC2018-094773, PGC2018-102021, SEV-2016-0588, SEV-2016-0597, and MDM-2015-0509, some of which include ERDF funds from the European Union. IFAE is partially funded by the CERCA program of the Generalitat de Catalunya.
Research leading to these results has received funding from the European Research
Council under the European Union's Seventh Framework Program (FP7/2007-2013) including ERC grant agreements 240672, 291329, and 306478.
We  acknowledge support from the Brazilian Instituto Nacional de Ci\^encia
e Tecnologia (INCT) do e-Universo (CNPq grant 465376/2014-2).

This research used resources of the National Energy Research Scientific Computing Center (NERSC), a U.S. Department of Energy Office of Science User Facility located at Lawrence Berkeley National Laboratory, operated under Contract No. DE-AC02-05CH11231 using NERSC award HEP-ERCAP0023923.
This manuscript has been authored by Fermi Research Alliance, LLC under Contract No. DE-AC02-07CH11359 with the U.S. Department of Energy, Office of Science, Office of High Energy Physics.
} 

\textit{Facilities: }
CTIO:4m, AAT, Gemini:Gillett (GMOS-N), Gemini:South (GMOS-S), Keck:I (LRIS), Keck:II (DEIMOS), Magellan:Baade (IMACS), Magellan:Clay (LDSS3, MagE), SALT

\textit{Software: }
\texttt{numpy} \citep{numpy}, 
\texttt{CAMB} \citep{camb}, 
\texttt{matplotlib} \citep{matplotlib}, 
\texttt{scipy} \citep{scipy}, 
\texttt{SNANA} \citep{Snana}, 
\texttt{Pippin} \citep{Pippin}, 
\texttt{Polychord} \citep{Polychord}, 
\texttt{Anesthetic} \citep{Anesthetic},
\texttt{PlikLite} Python implementation \citep{prince19},
\texttt{HMCODE2020} \citep{Mead2020},
\texttt{TurboGL} \citep{Kainulainen2009}

\section*{Data Availability}
The data and Python code used to generate the results and plots in this paper are available on reasonable request from the authors. The lensing pdfs used will be made available upon publication at \url{https://github.com/paulshah/SNLensing}.
 



\bibliographystyle{mnras}
\bibliography{DEScitations} 



\appendix
\section{Systematic error estimation}
\label{sec:validation}
\subsection{The lensing likelihood}
\label{sec:likelihood_validation}

Although we have given plausible reasons for our assignment of the likelihood, doubts may still arise on its validity. One may worry about the ability of the sin/arcsin distribution to accurately represent the non-Gaussianity of intrinsic effects and residual contamination by non SN Ia. Also, bias corrections adjust the mean of the residual distribution in bins to account for selection effects. 
\par
The Bayesian framework of estimated credible intervals allows us to exploit rigorous theorems to establish whether a posterior or specified credible interval is accurate. We note that while the homogeneous Gaussian SN Ia likelihood $\mathcal{L}_G$ has been extensively tested against simulations \citep[see for example][]{Camilleri2024}, even this commonly-used likelihood remains (in the sense we explain below) unproven. 
\par
Constraints on parameters are normally quoted as credible intervals, defined to be the range in which the integrated probability mass takes the specified value. A credible interval is located arbitrarily: common choices are to center it on the mean, the median, or define it as the region of highest probability density (such that the size of the interval in parameter space is minimized). An inference pipeline such as the one we have described in this paper is a process to assign an \textit{estimated} posterior distribution $\hat{p}(\theta | d)$ to model parameters $\theta$ given the data $d$. Intuitively, one may anticipate that if the estimated posterior correctly reproduces \textit{all} of the credible intervals (with arbitrary location) of the true distribution $p$, then the posterior estimation is correct and $\hat{p} = p$. This was made rigorous in \cite{Lemos2023}, who also give efficient methods for the computation. 
\par 
Their method can be summarised as follows:
\begin{itemize}
    \item choose a credibile level $1-\alpha$, a proposal function $g$ for the centre of the credibile interval $\theta_r$, and a distance metric on the space of model parameters $s(\theta_1, \theta_2)$
    \item generate $i = 1 \ldots N_{\rm sim}$ simulations of data vectors $d_i$ drawing truth model parameters $\theta_{i}^{*}$ from the prior 
    \item for each simulation $i$, construct the estimated posterior $\hat{p}_i$ and draw $j = 1 \ldots N_{\rm sample}$ samples $\theta_{ij}$ from it
    \item the coverage probability for each simulation is $f_i = (1/N_{\rm sample}) \sum_j \mathcal{I} [s(\theta_{ij}, \theta_r) < s(\theta_{i, *}, \theta_r)]$ where $\mathcal{I}$ is the indicator function
    \item the expected coverage probability (ECP) for the estimated posterior method is then ECP $ = (1/N_{\rm sim}) \sum_i \mathcal{I} (f_i < 1 -\alpha)$
\end{itemize}
If the ECP = $1-\alpha$ for all $\alpha$ and proposal functions $g$, then Theorem 3 of \cite{Lemos2023} states that $\hat{p} = p$. In our context, this procedure represents a comparison of our analytic likelihood to an implicitly-defined simulation-based likelihood that is deemed to be the truth. 

\begin{figure}
\label{fig:posterior_validation}
\centering 
\includegraphics[width=0.9\columnwidth]{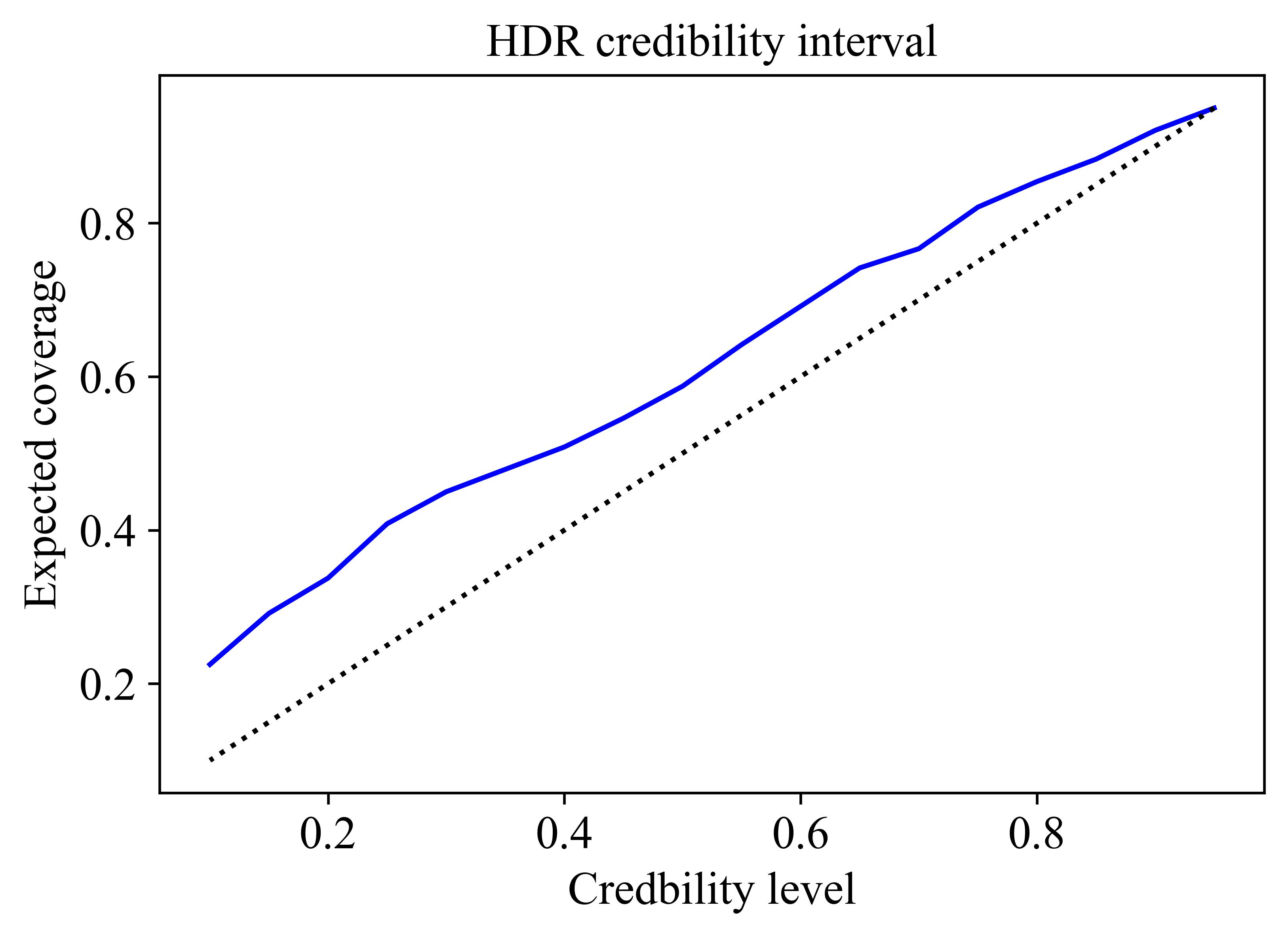}
\includegraphics[width=0.9\columnwidth]{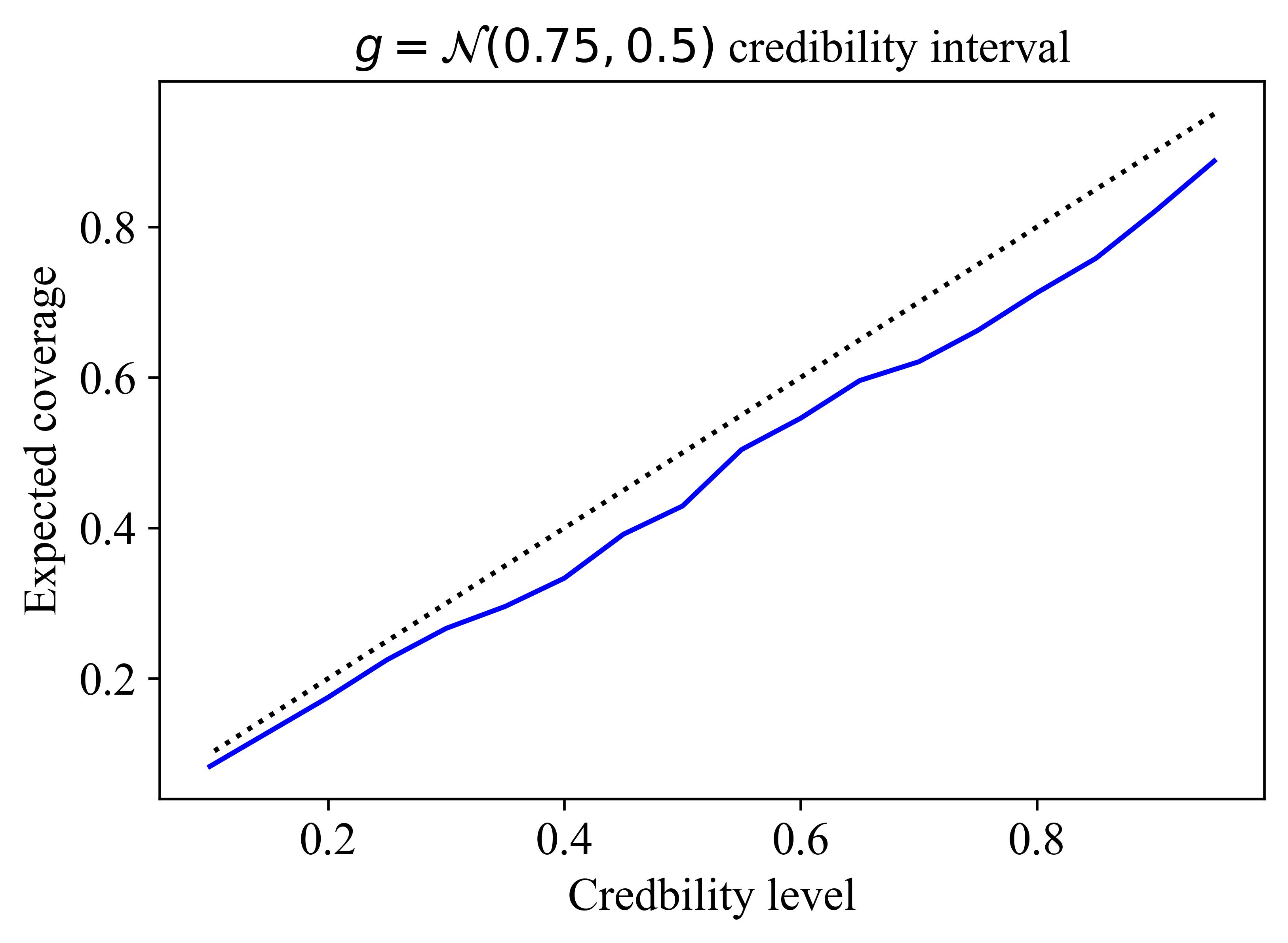}
\includegraphics[width=0.9\columnwidth]{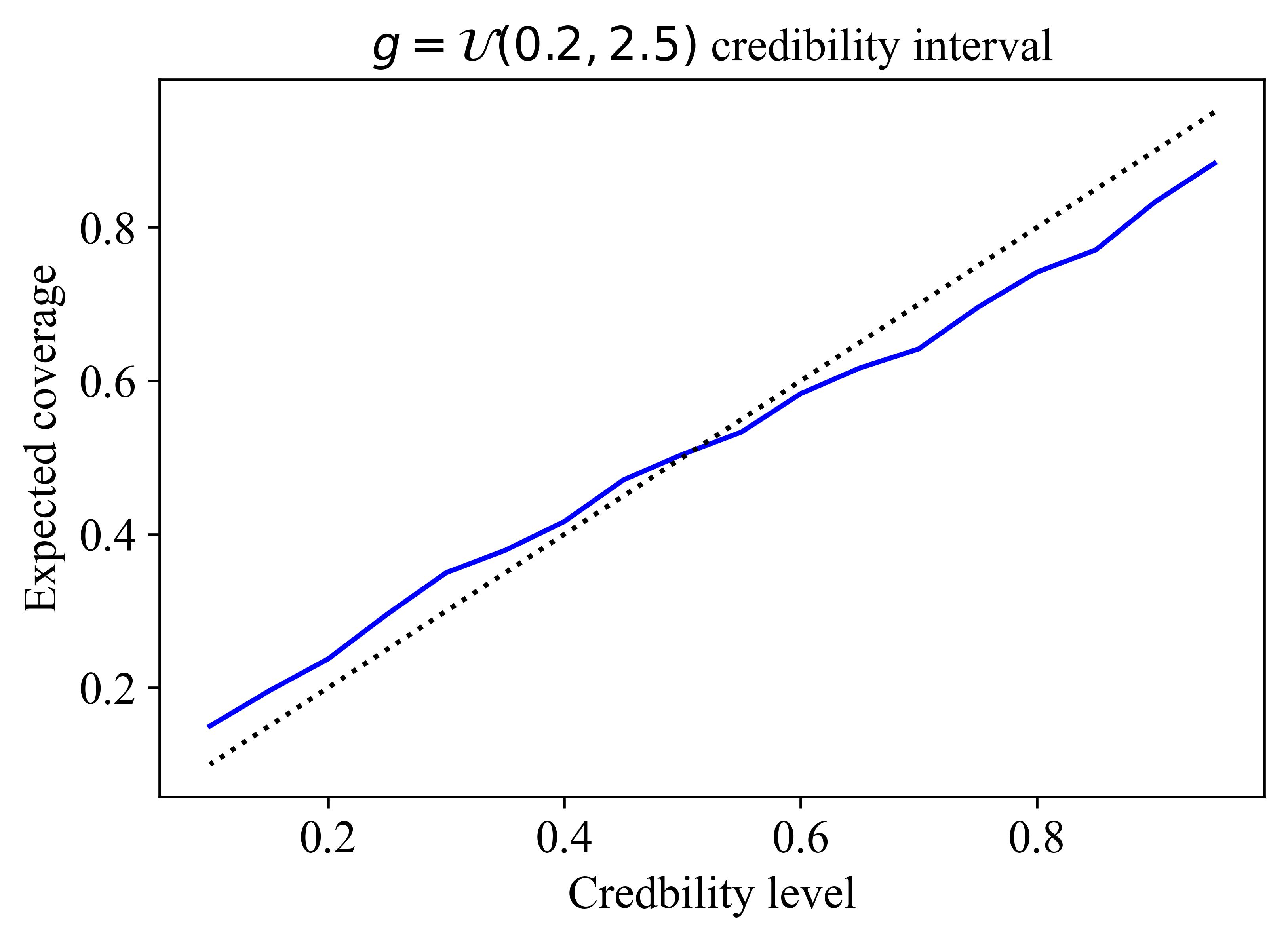}
\caption{Graphs of the expected coverage probability (ECP) (blue) against the specified credible interval for three choices of the credible interval location proposal function $g$. The x-axis is the integrated probability mass within the credible interval, and the y-axis are the results of the validation procedure outlined in the text above. The correct case is the black dashed line. An under-confident likelihood would result in the blue line above the black dashed line: the truthful credible interval level $1-\alpha$ is larger than the desired input. Conversely, an over-confident likelihood would lie under. A biased likelihood would mix under-confidence in some intervals with over-confidence elsewhere. \textit{Upper panel.} The estimation of the highest-density credible interval is under-confident by between $0\% - 20\%$. \textit{Middle panel.} The intervals concentrated around the median are over-confident by up to $\sim 7\%$. \textit{Lower panel.} The intervals randomly located across the prior show the likelihood is biased to a similar degree, confirming the results of the upper two panels.}
\end{figure}

We adopt three choices for $g$. We validate the regions of highest probability density (HPD) of the 1-d marginalised distribution of $A_{\rm mod}$, a uniform $g = \mathcal{U}(0.2,2.5)$, and a normal $g = \mathcal{N}(0.75,0.5)$ (i.e. in the vicinity of the median). The distance metric is $s(\theta_1, \theta_2) = | \theta_1 - \theta_2 |$. We generate 240 simulations using the \texttt{SNANA} software \citep{Snana} with pipeline coordinator \texttt{PIPPIN} \citep{Pippin}, 10 for each uniformly spaced $A_{\rm mod}$ value between the limits of the prior $A_{\rm mod} \in (0.2, 2.5)$. The simulations mimic the observing conditions and selection functions of the DES SN Ia survey, and are processed using the same pipeline as used for the real data, including adjustments for bias corrections. For each simulation, we sample $10^4$ times from the posterior. 
\par
Figure \ref{fig:posterior_validation}1 shows the results. Although the HPD credible intervals (our preferred statistic) are all under-confident, the likelihood shows evidence of inaccuracy at the $\sim 10\%$ level, and credible intervals in the vicinity of the median are somewhat over-confident. We can approximately convert this inaccuracy in the credible interval to a systematic error in $A_{\rm mod}$ in the following way : for a given location and desired credible interval $1-\alpha$, interpolate the ECP until it matches $1-\alpha$, and then re-evaluate the statistic at the credible interval matching this ECP. For example, for the 68\% HDR constraint, we see the ECP (the y-axis on the upper panel of Figure \ref{fig:posterior_validation}) that matches this corresponds to a $\sim 59\%$ credible interval (the x-axis) for our estimated posterior. Re-evaluating the posterior with this revised interval leads to a constraint of $A_{\rm mod} < 0.90$, indicating our result was conservative by $\Delta A_{\rm mod} \sim 0.19$. Evaluations of other intervals at such intermediate levels of credibility lead to similar, or smaller, systematic error estimates. 
\par
We therefore assign $\Delta A_{\rm mod} = 0.19$ as the systematic error due to our likelihood assignment.

\subsection{Cosmological priors}
\label{sec:prior_validation}
It has been noted in the literature that assigning CMB priors to the amplitude of linear fluctuations can result in galaxy shear analyses preferring greater power spectrum suppression compared to adopting a wider prior (see Figure 6 of \citet{GarciaGarcia2024} and Figure 11 of \citet{Terasawa2024}). This is likely due the combined influence of linear and intermediate scales on galaxy shear data, which we do not expect to occur for SN Ia lensing constraints (see Figure \ref{fig:sigm_by_kmax}). To check this, we have repeated our analysis using priors derived from combining the DES 3x2pt \citep{DES3x2} and baryon acoustic oscillations derived from the Sloan Digital Sky Survey (SDSS) \citep[as summarized in][]{Alam2021} (but retaining the Planck constraint on $n_s$). Note, \textit{we do not alter the reference benchmark where $A_{\rm mod}=1$ refers to fluctuations on all scales being consistent with the CMB in dark-matter only $\Lambda$CDM}. This would represent a change of our baseline unit rather than any physical effect, and inhibit comparison of the values derived here to each other or the literature.
\par 
The change of prior affects our $A_{\rm mod}$ mean value at the level of $\Delta A_{\rm mod} \sim 0.01$. This is confirmed by Figure \ref{fig:snbao_triangle}, as the posteriors for large and small scale fluctuation amplitudes are largely independent. Additionally varying $n_s$ within the constraints allowed by the CMB results in a similar very small variation. We do not consider variations in the sum of neutrino masses as this is one of the physical effects expected to be captured in the parameter $A_{\rm mod}$.

\subsection{Variation between TurboGL and power spectrum emulators}
\label{sec:turbogl_validation}
We have used \texttt{TurboGL} to construct our lensing pdf, and this forms our benchmark with respect to our $A_{\rm mod, S}$ result is obtained. Nevertheless, for the purposes of comparing with literature results, it is worthwhile to examine differences in lensing predictions from \texttt{TurboGL} to power spectrum emulators like \texttt{HMCODE2020}.
\par
Following the arguments of \citet{Marra2013}, we examine the second moment of the lensing pdf. This directly relates to the $A_{\rm mod}$ parameter, and the shape of the lensing pdf is broadly universal, so the higher moments that drive our constraints scale in proportion to the standard deviation. We assign error budgets to the following potential inaccuracies : 1) the use of linearisation in weak lensing 2) the halo mass function calibration of \citet{Sheth2001} 3) the halo concentration relation of \citet{Zhao2009} 4) uncertainties in the emulation of the power spectrum. 
\par
\begin{figure}
    \centering
    \includegraphics[width=\columnwidth]{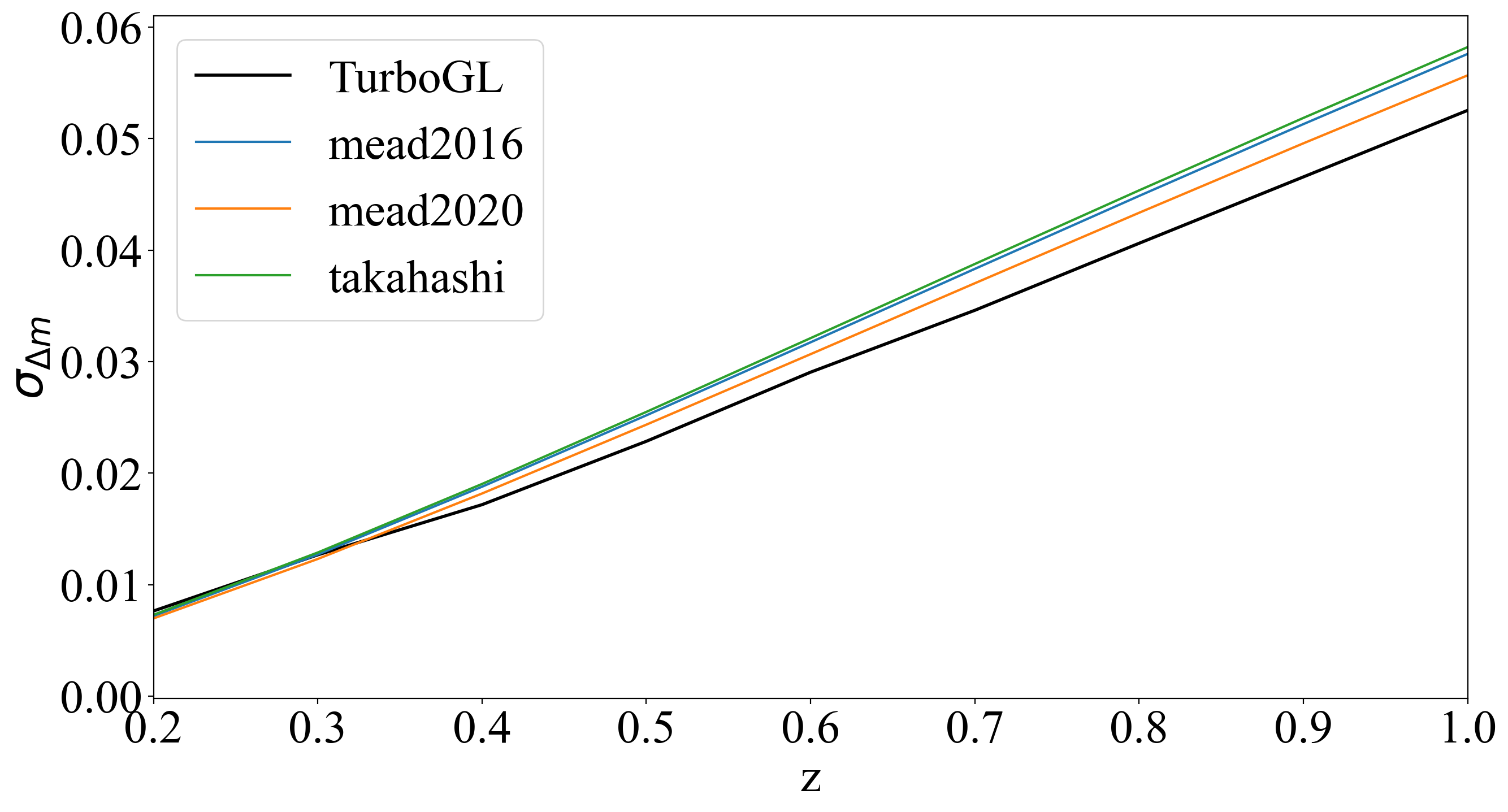} 
    \caption{The standard deviation of the lensing pdf (in magnitudes) by redshift. The literature models are as described in \citet{Mead2020, Mead2016} and \citet{Takahashi2012} respectively (coloured lines), and the calculation of $\sigma_{\Delta m}$ is made with Equation \ref{eq:frieman} with $k_{\rm max} = 10^3$. We compare this to the pdf we calculate using \texttt{TurboGL} (black line).}
    \label{fig:sigmalens_by_z}
\end{figure}

The first three were determined in \citet{Marra2013} as 5\%, 3\% and 3\% respectively and we adopt these estimates. For the last, we plot the variation in $\sigma_{\Delta m}$ (as determined from Equation \ref{eq:frieman} for $k_{\rm max} =10^3$) for the power spectrum models of \cite{Mead2020, Mead2016, Takahashi2012} alongside our results derived from \texttt{TurboGL} in Figure \ref{fig:sigmalens_by_z}. We adopt the median variation of 6\% as our systematic error for this component. Adding these errors in quadrature gives a total of 9\%.

\subsection{Summary}
To recap, we estimate systematic errors of $0.19$ for the likelihood, $0.01$ for cosmological priors, and $0.09$ for our lensing pdf. These add in quadrature to a total systematic error of $\Delta A_{\rm mod} = 0.21$. 

\noindent \\
$^{1}$ Department of Physics \& Astronomy, University College London, Gower Street, London, WC1E 6BT, UK\\
$^{2}$ School of Mathematics and Physics, University of Queensland,  Brisbane, QLD 4072, Australia\\
$^{3}$ Department of Physics, University of Oxford, Denys Wilkinson Building, Keble Road, Oxford OX1 3RH, UK\\
$^{4}$ The Research School of Astronomy and Astrophysics, Australian National University, ACT 2601, Australia\\
$^{5}$ Center for Astrophysics $\vert$ Harvard \& Smithsonian, 60 Garden Street, Cambridge, MA 02138, USA\\
$^{6}$ Institut d'Estudis Espacials de Catalunya (IEEC), 08034 Barcelona, Spain\\
$^{7}$ Institute of Space Sciences (ICE, CSIC),  Campus UAB, Carrer de Can Magrans, s/n,  08193 Barcelona, Spain\\
$^{8}$ SLAC National Accelerator Laboratory, Menlo Park, CA 94025, USA\\
$^{9}$ Department of Physics, University of Michigan, Ann Arbor, MI 48109, USA\\
$^{10}$ Department of Physics and Astronomy, University of Pennsylvania, Philadelphia, PA 19104, USA\\
$^{11}$ Centre for Gravitational Astrophysics, College of Science, The Australian National University, ACT 2601, Australia\\
$^{12}$ Centre for Astrophysics \& Supercomputing, Swinburne University of Technology, Victoria 3122, Australia\\
$^{13}$ School of Physics and Astronomy, University of Southampton,  Southampton, SO17 1BJ, UK\\
$^{14}$ Fermi National Accelerator Laboratory, P. O. Box 500, Batavia, IL 60510, USA\\
$^{15}$ Laborat\'orio Interinstitucional de e-Astronomia - LIneA, Rua Gal. Jos\'e Cristino 77, Rio de Janeiro, RJ - 20921-400, Brazil\\
$^{16}$ Department of Physics, Northeastern University, Boston, MA 02115, USA\\
$^{17}$ Instituto de Astrofisica de Canarias, E-38205 La Laguna, Tenerife, Spain\\
$^{18}$ Universidad de La Laguna, Dpto. Astrofísica, E-38206 La Laguna, Tenerife, Spain\\
$^{19}$ Institut de F\'{\i}sica d'Altes Energies (IFAE), The Barcelona Institute of Science and Technology, Campus UAB, 08193 Bellaterra (Barcelona) Spain\\
$^{20}$ Jodrell Bank Center for Astrophysics, School of Physics and Astronomy, University of Manchester, Oxford Road, Manchester, M13 9PL, UK\\
$^{21}$ University of Nottingham, School of Physics and Astronomy, Nottingham NG7 2RD, UK\\
$^{22}$ Hamburger Sternwarte, Universit\"{a}t Hamburg, Gojenbergsweg 112, 21029 Hamburg, Germany\\
$^{23}$ Department of Physics, IIT Hyderabad, Kandi, Telangana 502285, India\\
$^{24}$ California Institute of Technology, 1200 East California Blvd, MC 249-17, Pasadena, CA 91125, USA\\
$^{25}$ Institute of Theoretical Astrophysics, University of Oslo. P.O. Box 1029 Blindern, NO-0315 Oslo, Norway\\
$^{26}$ Kavli Institute for Cosmological Physics, University of Chicago, Chicago, IL 60637, USA\\
$^{27}$ Instituto de Fisica Teorica UAM/CSIC, Universidad Autonoma de Madrid, 28049 Madrid, Spain\\
$^{28}$ Institute of Cosmology and Gravitation, University of Portsmouth, Portsmouth, PO1 3FX, UK\\
$^{29}$ University Observatory, Faculty of Physics, Ludwig-Maximilians-Universit\"at, Scheinerstr. 1, 81679 Munich, Germany\\
$^{30}$ Center for Astrophysical Surveys, National Center for Supercomputing Applications, 1205 West Clark St., Urbana, IL 61801, USA\\
$^{31}$ Department of Astronomy, University of Illinois at Urbana-Champaign, 1002 W. Green Street, Urbana, IL 61801, USA\\
$^{32}$ Santa Cruz Institute for Particle Physics, Santa Cruz, CA 95064, USA\\
$^{33}$ Center for Cosmology and Astro-Particle Physics, The Ohio State University, Columbus, OH 43210, USA\\
$^{34}$ Department of Physics, The Ohio State University, Columbus, OH 43210, USA\\
$^{35}$ Jet Propulsion Laboratory, California Institute of Technology, 4800 Oak Grove Dr., Pasadena, CA 91109, USA\\
$^{36}$ George P. and Cynthia Woods Mitchell Institute for Fundamental Physics and Astronomy, and Department of Physics and Astronomy, Texas A\&M University, College Station, TX 77843,  USA\\
$^{37}$ LPSC Grenoble - 53, Avenue des Martyrs 38026 Grenoble, France\\
$^{38}$ Instituci\'o Catalana de Recerca i Estudis Avan\c{c}ats, E-08010 Barcelona, Spain\\
$^{39}$ Department of Physics, Carnegie Mellon University, Pittsburgh, Pennsylvania 15312, USA\\
$^{40}$ Observat\'orio Nacional, Rua Gal. Jos\'e Cristino 77, Rio de Janeiro, RJ - 20921-400, Brazil\\
$^{41}$ Kavli Institute for Particle Astrophysics \& Cosmology, P. O. Box 2450, Stanford University, Stanford, CA 94305, USA\\
$^{42}$ Centro de Investigaciones Energ\'eticas, Medioambientales y Tecnol\'ogicas (CIEMAT), Madrid, Spain\\
$^{43}$ Ruhr University Bochum, Faculty of Physics and Astronomy, Astronomical Institute, German Centre for Cosmological Lensing, 44780 Bochum, Germany\\
$^{44}$ Physics Department, Lancaster University, Lancaster, LA1 4YB, UK\\
$^{45}$ Computer Science and Mathematics Division, Oak Ridge National Laboratory, Oak Ridge, TN 37831\\
$^{46}$ Department of Astronomy, University of California, Berkeley,  501 Campbell Hall, Berkeley, CA 94720, USA\\
$^{47}$ Lawrence Berkeley National Laboratory, 1 Cyclotron Road, Berkeley, CA 94720, USA\\


\bsp	
\label{lastpage}
\end{document}